\documentclass[letter,10pt,conference]{IEEEtran}

 \usepackage[bookmarks=false]{hyperref}
\ifCLASSINFOpdf
\else
\fi
\usepackage{relsize}

\usepackage{centernot}
\usepackage{cite}
\usepackage{amsfonts}
\usepackage{mathrsfs}
\usepackage{bbm}
\usepackage{pifont}
\usepackage{amsfonts}
\usepackage{amsmath, amsthm}
\usepackage[top=.6in,bottom=.7in,left=.61in,right=.61in]{geometry}

\usepackage{tabularx}
\usepackage{listings}
\usepackage{graphicx}
\usepackage{float}
\usepackage{amssymb}
\usepackage{array}
\usepackage{fancyhdr}
\usepackage{cases}
 \usepackage{supertabular}
\usepackage{url}
\usepackage{fancyhdr}

\usepackage{psfrag}
\usepackage{color}
\usepackage{bbding}

\newcommand {\C} {{\rm I\kern-5.5pt C}}

\def\centerhack#1{\hbox to 0pt{\hss\footnotesize #1\hss}}
\def\centerhackn#1{\hbox to 0pt{\hss #1\hss}}
\def\dchack#1{\vbox to 0pt{\vss{\hbox to 0pt{\hss#1\hss}}\vss}}

\newtheorem{lem}{Lemma}
\newtheorem{thm}{Theorem}

\newtheorem*{proposition1.1}{Proposition 1.1}
\newtheorem*{proposition1.2}{Proposition 1.2}
\newtheorem*{proposition1.3}{Proposition 1.3}
\newtheorem*{proposition2.1}{Proposition 2.1}
\newtheorem*{proposition2.2}{Proposition 2.2}
\renewcommand\baselinestretch{.97}

\begin{document}

%

\title{Monotone Increasing Properties and Their Phase Transitions in Uniform Random Intersection Graphs}






 \author{ \IEEEauthorblockN{Jun Zhao}
\IEEEauthorblockA{CyLab and Dept.
of ECE \\
Carnegie Mellon University \\
{\tt junzhao@cmu.edu}} \and \IEEEauthorblockN{Osman Ya\u{g}an}
\IEEEauthorblockA{CyLab and Dept.
of ECE\\
Carnegie Mellon University \\
{\tt oyagan@ece.cmu.edu}} \and \IEEEauthorblockN{Virgil Gligor}
\IEEEauthorblockA{CyLab and Dept.
of ECE \\
Carnegie Mellon University \\
{\tt gligor@cmu.edu}}}

\maketitle \thispagestyle{plain} \pagestyle{plain}

%
%



\maketitle


 \begin{abstract}

 Uniform random intersection graphs have received much interest and been used in diverse applications.
 A uniform random intersection graph with $n$ nodes is constructed as follows: each node selects a set of $K_n$ different items uniformly at random from the same pool of $P_n$ distinct items, and two nodes establish an undirected edge in between if and only if they share at least one item. For such graph denoted by $G(n, K_n, P_n)$, we present the following results in this paper. First, we provide an exact analysis on the probabilities of
 $G(n, K_n, P_n)$ having a perfect matching and having a Hamilton cycle respectively, under $P_n = \omega\big(n (\ln n)^5\big)$ (all asymptotic notation are understood with $n \to \infty$). The analysis reveals that just like ($k$-)connectivity shown in prior work, for both properties of perfect matching containment and Hamilton cycle containment,  $G(n, K_n, P_n)$ also exhibits phase transitions: for each property above, as $K_n$ increases, the limit of the probability that $G(n, K_n, P_n)$ has the property increases from $0$ to $1$. Second, we compute the phase transition widths of $G(n, K_n, P_n)$ for $k$-connectivity, perfect matching containment, and Hamilton cycle containment, respectively. For a graph property $\mathcal{I}$ and a positive constant $\epsilon < \frac{1}{2}$, with the phase transition width $d_n(\mathcal{I}, \epsilon)$ defined as the non-negative difference between the minimal $K_n$ ensuring $G(n, K_n, P_n)$ having property $\mathcal{I}$ with probability at least $1-\epsilon$, and the minimal $K_n$ ensuring $G(n, K_n, P_n)$ having property $\mathcal{I}$ with probability at least $\epsilon$, we show for any positive constant $\epsilon < \frac{1}{2}$ and any positive constant integer $k$ that:\\
(i) If $P_n = \Omega(n)$ and $P_n = o(n\ln n)$, $d_n(\text{$k$-connectivity}, \epsilon)$ equals either $0$ or $1$ for each $n$ sufficiently large.\\
(ii) If $P_n = \Theta(n\ln n)$, then $d_n(\text{$k$-connectivity}, \epsilon)=\Theta(1)$.\\
(iii) If $P_n = \omega(n\ln n)$, then $d_n(\text{$k$-connectivity}, \epsilon)=\omega(1)$.\\
(iv) If $P_n = \omega\big(n (\ln n)^5\big)$, $d_n(\text{perfect matching containment}, \epsilon)$ and $d_n(\text{Hamilton cycle containment}, \epsilon)$ can both be written as $\omega(1)$.

\end{abstract}
%
%

 
 \section{Introduction}

   Uniform random intersection graphs have received much attention and been used in various applications \cite{r1,Perfectmatchings,NikoletseasHM,r4,ryb3,zz,2013arXiv1301.0466R,yagan,ZhaoYaganGligor,ZhaoCDC,ANALCO}. A uniform random intersection graph with $n$ nodes is defined as follows: each node picks a set of $K_n$ different items \emph{uniformly at random} from the same pool of $P_n$ distinct items, and an \emph{undirected} edge is put between
   any two nodes which share at least one item. We will denote a uniform random intersection graph by $G(n, K_n, P_n)$.
    Uniform random intersection graphs belong to a wider class of graphs called random intersection graphs in which each node selects some items in a random manner and any two nodes have an undirected edge upon sharing a certain number of items \cite{zz,2013arXiv1301.0466R,yagan,ZhaoYaganGligor,ZhaoCDC,ANALCO}. 
   
   Uniform random intersection graphs are also referred to as random key graphs due to their applications to the Eschenauer--Gligor key predistribution scheme \cite{virgil}, which is a recognized approach to ensure secure
communications in wireless sensor networks. In the Eschenauer--Gligor scheme for a wireless sensor network with $n$ sensors, 
before deployment,
 each sensor is assigned a set of $K_n$ distinct cryptographic keys selected uniformly at random from the same key pool containing $P_n$ different keys. After deployment, two sensors establish secure communication if
and only if they have at least one common key. Clearly the induced topology is a uniform random intersection graph. In addition to secure sensor networks, uniform random intersection graphs have been used for 
 recommender systems \cite{r4}, social networks \cite{ZhaoYaganGligor}, and circuit design \cite{2013arXiv1301.0466R}.

In this paper, we study monotone increasing properties and their phase transitions in uniform random intersection graphs. The studied properties include $k$-connectivity, perfect matching containment, and Hamilton cycle containment. First, $k$-connectivity means that each pair of nodes has at least $k$
internally node-disjoint path(s) between them \cite{ZhaoYaganGligor}. Second, a perfect matching in a graph with an even number of nodes means a
matching covering all nodes, where a matching in a graph is a set of edges without common nodes  \cite{Perfectmatchings}.
    We use the generalized notion of perfect matching: for a graph with an odd number of nodes, a perfect matching
   means a
matching covering all nodes except one \cite{Perfectmatchings}. Finally, a Hamiltonian cycle in a graph is a closed loop that visits each node once \cite{EfthymiouaHM}.

  The above studied properties of uniform random intersection graphs have various applications. In the use of uniform random intersection graphs for secure wireless sensor networks \cite{virgil}, $k$-connectivity enables multi-path routing and load balancing,
  and is useful for consensus \cite{ZhaoYaganGligor};
   perfect matchings have been used for the analysis of wireless information flow \cite{Amaudruz2009} and network coverage \cite{Wang20091427}, 
  and the optimal allocation of rate and power \cite{4407772}; and Hamilton cycles have been used for cyclic routing which along with distributed optimization achieves efficient in-network data processing \cite{1413472}.

We make the following contributions in this paper: i) we provide an exact analysis on the probabilities of
 a uniform random intersection graph $G(n, K_n, P_n)$ having a perfect matching and having a Hamilton cycle respectively, and ii) we compute the phase transition widths of $G(n, K_n, P_n)$ for $k$-connectivity, perfect matching containment, and Hamilton cycle containment, respectively. 
   Note that
when we say a graph has a perfect matching (respectively, Hamilton cycle), we mean the graph has \emph{at least one} perfect matching (respectively, Hamilton cycle).



    The rest of the paper is organized as follows. Section \ref{sec:main:res} presents the main results as
theorems. Then, we introduce several auxiliary lemmas in
Section \ref{sec:factlem}, before establishing the theorems in
Section \ref{sec:thmprf:kcon}. Section
\ref{related} reviews related work, and Section \ref{sec:Conclusion}
concludes the paper. The Appendix
details the proofs of the lemmas.

   \section{The Main Results} \label{sec:main:res}
   
  We present the main results in Theorems \ref{thm:PM}--\ref{thm:PHW}.  We use the standard asymptotic notation $\Omega(\cdot), \omega(\cdot), O(\cdot),
o(\cdot),\Theta(\cdot)$. All asymptotics and limits are taken with $n \to \infty$. Also, $\mathbb{P}[\cdot]$
denotes an event \vspace{-1pt} probability. An event happens \emph{with high probability} if its probability converges to $1$ as $n\to\infty$.

   \subsection{An exact analysis on perfect matching containment and Hamilton cycle containment}
   
   \begin{thm} \label{thm:PM}
For a uniform random intersection graph
$G(n,K_n,P_n)$, if there is a sequence $\alpha_n$ with $\lim_{n \to \infty}{\alpha_n} \in [-\infty, \infty]$
such that
\begin{align}
 \frac{{K_n}^2}{P_n} & = \frac{\ln  n   +
 {\alpha_n}}{n}, \label{thm:PM:pe}
\end{align}
then under $P_n = \omega\big(n (\ln n)^5\big)$, it holds that
 \begin{align}
& \lim_{n \to \infty}  \mathbb{P}[\hspace{2pt}G(n, K_n, P_n)\text{ has a perfect matching.}\hspace{2pt}] \nonumber \\ &   \hspace{-1pt}=\hspace{-1.5pt} e^{- e^{-\lim\limits_{n \to \infty}{\alpha_n}}} \hspace{-1.5pt}=\hspace{-1.5pt}  \begin{cases} 0,&\text{\hspace{-4pt}if  }\lim_{n \to \infty}{\alpha_n} \hspace{-1.5pt}=\hspace{-1.5pt}- \infty, \\
1,&\text{\hspace{-4pt}if  }\lim_{n \to \infty}{\alpha_n} \hspace{-1.5pt}=\hspace{-1.5pt} \infty, \\
e^{-e^{- \alpha^{*}}},&\text{\hspace{-4pt}if  }\lim_{n \to \infty}{\alpha_n} \hspace{-1.5pt}=\hspace{-1.5pt} \alpha^{*}\hspace{-1.5pt}\in\hspace{-1.5pt} (-\infty, \hspace{-1.5pt}\infty). \end{cases}\nonumber
 \end{align}
\end{thm}

%
%

   \begin{thm} \label{thm:HC}
For a uniform random intersection graph
$G(n,K_n,P_n)$, if there is a sequence $\beta_n$ with $\lim_{n \to \infty}{\beta_n} \in [-\infty, \infty]$
such that
\begin{align}
 \frac{{K_n}^2}{P_n} & = \frac{\ln  n +  \ln \ln n +
 {\beta_n}}{n}, \label{thm:HC:pe}
\end{align}
then under $P_n = \omega\big(n (\ln n)^5\big)$, it holds that
 \begin{align}
& \lim_{n \to \infty}  \mathbb{P}[\hspace{2pt}G(n, K_n, P_n)\text{ has a Hamilton cycle.}\hspace{2pt}] \nonumber \\ &   \hspace{-1pt}=\hspace{-1.5pt} e^{- e^{-\lim\limits_{n \to \infty}{\beta_n}}} \hspace{-1.5pt}=\hspace{-1.5pt}  \begin{cases} 0,&\text{\hspace{-4pt}if  }\lim_{n \to \infty}{\beta_n} \hspace{-1.5pt}=\hspace{-1.5pt}- \infty, \\
1,&\text{\hspace{-4pt}if  }\lim_{n \to \infty}{\beta_n} \hspace{-1.5pt}=\hspace{-1.5pt} \infty, \\
e^{-e^{- \beta^{*}}},&\text{\hspace{-4pt}if  }\lim_{n \to \infty}{\beta_n} \hspace{-1.5pt}=\hspace{-1.5pt} \beta^{*}\hspace{-1.5pt}\in\hspace{-1.5pt} (-\infty, \hspace{-1.5pt}\infty). \end{cases}\nonumber
 \end{align}
\end{thm}

%

Theorems \ref{thm:PM} and \ref{thm:HC} show that uniform random intersection graphs exhibit phase transitions for perfect matching containment and Hamilton cycle containment.
By \cite[Lemma 8]{ZhaoYaganGligor}, the term $ \frac{{K_n}^2}{P_n}$ in (\ref{thm:PM:pe}) and (\ref{thm:HC:pe})
is an asymptotic value of the edge probability $1-\binom{P_n-K_n}{K_n}\big/\binom{P_n}{K_n}$ (i.e., the probability for the existence of an edge between two nodes). If we \vspace{1.5pt} replace $ \frac{{K_n}^2}{P_n}$ in (\ref{thm:PM:pe}) and (\ref{thm:HC:pe}) with the edge probability, Theorems \ref{thm:PM} and \ref{thm:HC}  still follow; see the explanations given in the proofs of Theorems \ref{thm:PM} and \ref{thm:HC}.

Note that a difference between a uniform random intersection graph and an Erd\H{o}s--R\'enyi graph \cite{erdosPF} which is constructed by assigning an edge between each pair of nodes independently with the same probability, is that not all edges in the former graph are independent, while in the latter graph all edges are mutually independent. Despite of this difference, the results above along with Lemmas \ref{lem:ER:PM} and \ref{lem:ER:HC} in Section \ref{sec:factlem} show that in both graphs,
common critical thresholds of the edge probability are $\frac{\ln  n }{n}$ for perfect matching containment, and $\frac{\ln  n +  \ln \ln n }{n}$ for Hamilton cycle containment.

   \subsection{Phase transition widths for $k$-connectivity, perfect matching containment and Hamilton cycle containment}

For each $n$, given $P_n$, the probability that $G(n, K_n, P_n)$ has a monotone increasing graph property $\mathcal{I}$ increases as $K_n$ increases \cite{zz}, so we define for a positive constant $\epsilon < \frac{1}{2}$ that\vspace{-2pt}
\begin{align}
K_n^{-}(\mathcal{I}, \epsilon) :=\vspace{-2pt} \min \bigg\{K_n \bigg| \begin{array}{l}G(n, K_n, P_n)\text{ has property }\mathcal{I}\\ \text{  with probability at least }\epsilon .\end{array}\bigg\} \label{KnI-minus}
\end{align}
and\vspace{-2pt}
\begin{align}
K_n^{+}(\mathcal{I}, \epsilon) :=\vspace{-2pt} \min \bigg\{K_n \bigg| \begin{array}{l}G(n, K_n, P_n)\text{ has property }\mathcal{I}\\ \text{  with probability at least }1-\epsilon .\end{array}\bigg\}.\label{KnI-plus}
\end{align}
The phase transition width $d_n(\mathcal{I}, \epsilon)$ of graph $G(n,K_n,P_n)$ for a monotone increasing graph property $\mathcal{I}$ and a positive constant $\epsilon < \frac{1}{2}$ is defined by\vspace{-2pt}
\begin{align}
d_n(\mathcal{I}, \epsilon)\vspace{-2pt} = K_n^{+}(\mathcal{I}, \epsilon) - K_n^{-}(\mathcal{I}, \epsilon). \label{phase-transition-width}
\end{align}

   \begin{thm} \label{thm:PHW}
   For a uniform random intersection graph
$G(n,K_n,P_n)$, 
 the following results hold for any positive constant $\epsilon < \frac{1}{2}$ and any positive constant integer $k$:\\
  (i) If $P_n = \Omega(n)$ and $P_n = o(n\ln n)$, $d_n(\text{$k$-connectivity}, \epsilon)$ equals either $0$ or $1$ for each $n$ sufficiently large.\\
  (ii) If $P_n = \Theta(n\ln n)$, then $d_n(\text{$k$-connectivity}, \epsilon)=\Theta(1)$.\\
  (iii) If $P_n = \omega(n\ln n)$, then $d_n(\text{$k$-connectivity}, \epsilon)=\omega(1)$.\\
  (iv) If $P_n \hspace{-.5pt}=\hspace{-.5pt} \omega\big(n (\ln n)^5\big)$, $d_n(\text{perfect matching containment}, \epsilon)$ and $d_n(\text{Hamilton cycle containment}, \epsilon)$ can both be written as $\omega(1)$. \vspace{-2pt}
   \end{thm}
   
   The result (i) in Theorem \ref{thm:PHW} above shows that under $P_n = \Omega(n)$ and $P_n = o(n\ln n)$, a uniform random intersection graph exhibits a sharp phase transition for $k$-connectivity for each $n$ sufficiently large: if $G(n,K_n,P_n)$ has a probability of at least $\epsilon$ to be $k$-connected, then $G(n,K_n+1,P_n)$ has a probability of at least $1-\epsilon$ to be $k$-connected, since $d_n(\text{$k$-connectivity}, \epsilon)$ is at most $1$. Note that $d_n(\text{$k$-connectivity}, \epsilon)$ could be $0$; e.g., for some $K_n^*$, if the probabilities of $G(n,K_n^*,P_n)$ and $G(n,K_n^*+1,P_n)$ being $k$-connected are less than $\epsilon$, and at least $1-\epsilon$, respectively, then $d_n(\text{$k$-connectivity}, \epsilon) = 0$ follows since $K_n^{-}(\text{$k$-connectivity}, \epsilon)$ and $K_n^{+}(\text{$k$-connectivity}, \epsilon)$ given by (\ref{KnI-minus}) and (\ref{KnI-plus}) both equal $K_n^*+1$ in this case.
   
 For a uniform random intersection graph
$G(n,K_n,P_n)$, the results on $k$-connectivity in Lemma \ref{thm:ZhaoCDCk} in the next section use $P_n = \Omega(n)$, while the results on perfect matching containment and Hamilton cycle containment in Theorems \ref{thm:PM} and \ref{thm:HC} rely on $P_n = \omega\big(n (\ln n)^5\big)$. Since Theorem \ref{thm:PHW} above is established from Lemma \ref{thm:ZhaoCDCk}, and Theorems \ref{thm:PM} and \ref{thm:HC},  then Theorem \ref{thm:PHW} has results for $k$-connectivity under $P_n = \Omega(n)$, and for perfect matching containment and Hamilton cycle containment under a narrower range of $P_n = \omega\big(n (\ln n)^5\big)$; i.e., for $P_n = \Omega(n)$ and $P_n = O\big(n (\ln n)^5\big)$, we have phase transition results 
 for $k$-connectivity, but not for perfect matching containment and Hamilton cycle containment. A future direction is to extend our results to a wider range of $P_n$.   
   \section{Auxiliary Lemmas} \label{sec:factlem}

 We provide a few lemmas used to establish the
theorems. The proofs of Lemmas 
\ref{thm:ZhaoCDCk} and \ref{cp_rig_er}  are given in the Appendix, while Lemmas \ref{lem-only-prove-lnlnn}, \ref{lem:ER:PM}, \ref{lem:ER:HC}, and \ref{lem:subseq} are results directly from prior work, and the proof of Lemma \ref{lem:subsequenc} is omitted since it is straightforward.

 Lemma 
\ref{thm:ZhaoCDCk} presents an exact analysis of $k$-connectivity in a uniform random intersection graph
$G(n,K_n,P_n)$. Since this lemma easily follows from our work \cite[Lemma 1]{ANALCO} and \cite[Theorem 1]{ZhaoCDC}, it is not emphasized in our contributions, but we still explain its proof in the Appendix for clarity.
 
 %

\begin{lem} \label{thm:ZhaoCDCk}

For a uniform random intersection graph
$G(n,K_n,P_n)$, if there is a sequence $\gamma_n$ with $\lim_{n \to \infty}{\gamma_n} \in [-\infty, \infty]$
such that
\begin{align}
 \frac{{K_n}^2}{P_n} & = \frac{\ln  n   + {(k-1)} \ln \ln n  +
 {\gamma_n}}{n}, \label{thm:pe:kgencdc}
\end{align}
then under $P_n = \Omega(n)$, it holds for a positive constant integer $k$ that\begin{align}
&  \lim_{n \to \infty}\mathbb{P} \left[\hspace{2pt}G(n,K_n,P_n)\textrm{
is $k$-connected}.\hspace{2pt}\right] \nonumber \\
 & = \lim_{n \to \infty}\mathbb{P} \left[\hspace{2pt}G(n,K_n,P_n)\textrm{
has a minimum degree at least $k$}.\hspace{2pt}\right]  \nonumber \\ & = e^{- \frac{e^{-\lim_{n \to \infty}{\gamma_n}}}{(k-1)!}}. \label{thm:pe:kgencdcres}
 \end{align}

\end{lem}

\begin{lem}[\hspace{0pt}{Our work \cite[Lemma 1]{ANALCO}}\hspace{0pt}] \label{lem-only-prove-lnlnn}

For a uniform random intersection graph
$G(n,K_n,P_n)$ under $P_n = \Omega(n)$ and (\ref{thm:pe:kgencdc}), the following results hold:

(i) If $\lim_{n \to \infty}\gamma_n = -\infty$, there exists graph $G(n,\widetilde{K_n},\widetilde{P_n})$ under $\widetilde{P_n} = \Omega(n)$ and
$\frac{{\widetilde{K_n}}^2}{\widetilde{P_n}} =  \frac{\ln  n + {(k-1)} \ln \ln n + {\widetilde{\gamma_n}}}{n}$ with $\lim_{n \to \infty}\widetilde{\gamma_n} = -\infty$ and $\widetilde{\gamma_n} = -O(\ln \ln n)$,
such that there exists a graph coupling\footnote{As used
by Rybarczyk \cite{zz,2013arXiv1301.0466R}, a coupling of two random graphs $G_1$ and
$G_2$ means a probability space on which random graphs $G_1'$ and
$G_2'$ are defined such that $G_1'$ and $G_2'$ have the same
distributions as $G_1$ and $G_2$, respectively. If $G_1'$ is a spanning subgraph
(resp., spanning supergraph) of $G_2'$, we say that under the coupling, $G_1$ is a spanning subgraph
(resp.,  spanning supergraph) of $G_2$, which yields that for any monotone increasing property $\mathcal {I}$, the probability of $G_1$ having $\mathcal {I}$ is at most (reap., at least) the probability of $G_2$ having $\mathcal {I}$.} under which
$G(n,K_n,P_n)$ is a spanning subgraph of $G(n,\widetilde{K_n},\widetilde{P_n})$.

(ii) If $\lim_{n \to \infty}\gamma_n = \infty$, there exists graph $G(n,\widehat{K_n},\widehat{P_n})$ under $\widehat{P_n} = \Omega(n)$ and 
$  \frac{{\widehat{K_n}}^{2}}{{\widehat{P_n}}}    = \frac{\ln  n + {(k-1)} \ln \ln n + {\widehat{\gamma_n}}}{n}$
with $\lim_{n \to \infty}\widehat{\gamma_n} = \infty$ and $\widehat{\gamma_n} = O(\ln \ln n)$,
such that there exists a graph coupling under which
$G(n,K_n,P_n)$ is a spanning supergraph of $G(n,\widehat{K_n},\widehat{P_n})$.

\end{lem}

\begin{lem} \label{cp_rig_er}

 If $K_n = \omega\big((\ln n)^3\big)$, $ \frac{K_n}{P_n} = o\left( \frac{1}{n\ln n} \right)$\vspace{1pt} and $ \frac{{K_n}^2}{P_n}  =
 o\left( \frac{1}{\ln n} \right)$, then there exists $s_n =
 \frac{{K_n}^2}{P_n} \cdot \left[1-
 o\left(\frac{1}{ \ln n}\right)\right]$ such that for any monotone increasing graph property $\mathcal{I}$, 
\begin{align}
  \mathbb{P}[\hspace{2pt}G(n,K_n,P_n)
  \textrm{ has }\mathcal{I}.
\hspace{2pt}]  
&  \geq
 \mathbb{P}[\hspace{2pt}G_{ER}(n,s_n)
 \textrm{ has }\mathcal{I}.\hspace{2pt} ] - o(1).  \label{cp_res_rig_er}
 \end{align}
 
 \end{lem} 
 
 Lemmas \ref{lem:ER:PM} and \ref{lem:ER:HC} below present results on Erd\H{o}s--R\'enyi graphs \cite{erdosPF}, where
 an $n$-node Erd\H{o}s--R\'enyi graph denoted by $G_{ER}(n,s_n)$ is constructed by assigning an edge between each pair of nodes independently with the same probability $s_n$.

\begin{lem}[\hspace{-.1pt}{\cite[Theorem 1]{erdosPF}}\hspace{0pt}] \label{lem:ER:PM}
For an Erd\H{o}s--R\'enyi graph $G_{ER}(n,s_n)$, if there is a sequence $\alpha_n$ with $\lim_{n \to \infty}{\alpha_n} \in [-\infty, \infty]$
such that $s_n  = \frac{\ln  n   +
 {\alpha_n}}{n}$, 
 then it holds that
 \begin{align}
 \lim_{n \to \infty}   \mathbb{P}[ G_{ER}(n,s_n)\text{ has a perfect matching.} ] &  = e^{- e^{-\lim\limits_{n \to \infty}{\alpha_n}}}. \nonumber
 \end{align}
 \end{lem}

 \begin{lem} [\hspace{-.1pt}{\cite[Theorem 1]{erdosHC}}\hspace{0pt}] \label{lem:ER:HC}
For an Erd\H{o}s--R\'enyi graph $G_{ER}(n,s_n)$, if there is a sequence $\beta_n$ with $\lim_{n \to \infty}{\beta_n} \in [-\infty, \infty]$
such that $s_n = \frac{\ln  n   + \ln \ln n +
 {\beta_n}}{n}$, 
 then it holds that
 \begin{align}
 \lim_{n \to \infty}   \mathbb{P}[ G_{ER}(n,s_n)\text{ has a Hamilton cycle.} ] &  = e^{- e^{-\lim\limits_{n \to \infty}{\beta_n}}}. \nonumber
 \end{align}
 \end{lem} 
%

 \begin{lem} [\hspace{-.1pt}{\cite[Lemma 1]{Xuan}}\hspace{0pt}]  \label{lem:subseq} 
Every sequence $a_n|_{n=1,2,\ldots}$ has a subsequence $a_{M_j} |_{j=1,2,\ldots}$ with $\lim_{j \to \infty} a_{M_j}  \in [-\infty, \infty]$, where $M_1<M_2<\ldots$ and $M_1, M_2, \ldots \in  \mathbb{N}$ ($\mathbb{N}$ is the set of 
 all positive integers).

\end{lem}
 
  \begin{lem}  \label{lem:subsequenc} 
In Theorems \ref{thm:PM} and \ref{thm:HC}, and Lemma \ref{thm:ZhaoCDCk}, if the conditions hold for just a subsequence of $\mathbb{N}$ instead of all $n \in \mathbb{N}$ , then the corresponding asymptotic results also hold for the subsequence.

\end{lem}
 
 \section{Establishing Theorems \ref{thm:PM}--\ref{thm:PHW}}
\label{sec:thmprf:kcon}

 \subsection{Proof of Theorem \ref{thm:PM}}

Let $\texttt{PerfMatch}$ be short for perfect matching.
Theorem \ref{thm:PM} follows once we prove the two inequalities below:
 \begin{align}
 \hspace{-1pt} \mathbb{P}[\hspace{.5pt}G(n, K_n, P_n)\text{ has a \texttt{PerfMatch}.}\hspace{.5pt}] & \hspace{-2pt} \leq \hspace{-2pt}e^{- e^{-\lim\limits_{n \to \infty}{\alpha_n}}} \hspace{-2pt} \cdot \hspace{-2pt} [1\hspace{-2pt}+\hspace{-2pt}o(1)]\label{PM-leq}
 \end{align}
 and
  \begin{align}
 \hspace{-1pt} \mathbb{P}[\hspace{.5pt}G(n, K_n, P_n)\text{ has a \texttt{PerfMatch}.}\hspace{.5pt}] & \hspace{-2pt} \geq \hspace{-2pt}e^{- e^{-\lim\limits_{n \to \infty}{\alpha_n}}} \hspace{-2pt}\cdot\hspace{-2pt} [1\hspace{-2pt}-\hspace{-2pt}o(1)].  \label{PM-geq}
 \end{align}
 
 (\ref{PM-leq}) clearly holds from Lemma \ref{thm:ZhaoCDCk} with $k=1$ and the fact that a necessary condition for a graph to contain a perfect matching is that the minimum degree is at least $1$ (i.e., there is no isolated node).
 
 Now we establish (\ref{PM-geq}). From Lemma \ref{lem-only-prove-lnlnn} and the fact that perfect matching containment is a monotone increasing graph property, we can introduce an auxiliary condition $|\alpha_n |= O(\ln \ln n)$. We will use Lemmas \ref{cp_rig_er} and \ref{lem:ER:PM} to prove (\ref{PM-geq}). We first show that the conditions of Lemma \ref{cp_rig_er} all hold given the conditions of Theorem \ref{thm:PM} and the auxiliary condition $|\alpha_n |= O(\ln \ln n)$. From $|\alpha_n |= O(\ln \ln n)$ and (\ref{thm:PM:pe}), it follows that
 \begin{align}
 \frac{{K_n}^2}{P_n} & = \frac{\ln  n}{n} \cdot [1 \pm o(1)]. \label{KnPnnlnn}
\end{align}
(\ref{KnPnnlnn}) implies $ \frac{{K_n}^2}{P_n}  =
 o\left( \frac{1}{\ln n} \right)$. Furthermore, (\ref{KnPnnlnn}) and $P_n = \omega\big(n (\ln n)^5\big)$ 
  together yield $K_n = \sqrt{ \frac{{K_n}^2}{P_n}  \cdot P_n }= \omega\big((\ln n)^3\big)$ and $ \frac{K_n}{P_n} =  \sqrt{\frac{{K_n}^2}{P_n}} / \sqrt{P_n} = o\big( \frac{1}{n(\ln n)^2} \big) = o\left( \frac{1}{n\ln n} \right)$. Then all conditions in Lemma \ref{cp_rig_er} hold. From (\ref{thm:PM:pe}) and (\ref{KnPnnlnn}), the sequence $s_n$ in Lemma \ref{cp_rig_er} satisfies 
 \begin{align}
s_n & =   \frac{\ln  n   +
 {\alpha_n}}{n} - \frac{\ln  n}{n} \cdot [1 \pm o(1)] \cdot  o\left(\frac{1}{ \ln n}\right) \nonumber \\ & =  \frac{\ln  n   +
 {\alpha_n}-o(1)}{n}  ,\label{KnPnnlnnres2}
\end{align}
which is used in Lemma \ref{lem:ER:PM} to induce
 \begin{align}
 \lim_{n \to \infty}   \mathbb{P}[ G_{ER}(n,s_n)\text{ has a \texttt{PerfMatch}.} ] &  = e^{- e^{-\lim\limits_{n \to \infty}{\alpha_n}}}. \label{KnPnnlnn3}
 \end{align}
 Then (\ref{PM-geq}) clearly follows from (\ref{KnPnnlnn3}) and Lemma \ref{cp_rig_er}.
 
We have established  Theorem \ref{thm:PM} by showing (\ref{PM-leq}) and (\ref{PM-geq}).

We now explain that Theorem \ref{thm:PM} still follows if $ \frac{{K_n}^2}{P_n}$ in (\ref{thm:PM:pe}) is replaced by the edge probability denoted by $q_n$; i.e., $q_n = \frac{\ln  n   +
 {\alpha_n}}{n}$. First, from \cite[Proof of Lemma 14]{ANALCO}, we can also introduce an auxiliary condition $|\alpha_n |= O(\ln \ln n)$, which along with $q_n = \frac{\ln  n   +
 {\alpha_n}}{n}$ yields $q_n = \frac{\ln  n}{n} \cdot [1 \pm o(1)]$. Then
from \cite[Lemma 8]{ZhaoYaganGligor}, it is straightforward to derive $ \frac{{K_n}^2}{P_n}  =  \frac{\ln  n   +
 {\alpha_n} \pm o(1)}{n} $. Thus Theorem \ref{thm:PM} clearly still follows.

  \subsection{Proof of Theorem \ref{thm:HC}}
  
  The proof of Theorem \ref{thm:HC} is similar to that of Theorem \ref{thm:PM}.
  
  Let $\texttt{HamiCycle}$ be short for Hamilton cycle.
Theorem \ref{thm:HC} follows once we prove the two inequalities below:
 \begin{align}
 \hspace{-1pt} \mathbb{P}[\hspace{.5pt}G(n, K_n, P_n)\text{ has a \texttt{HamiCycle}.}\hspace{.5pt}] & \hspace{-2pt} \leq \hspace{-2pt}e^{- e^{-\lim\limits_{n \to \infty}{\beta_n}}} \hspace{-2pt} \cdot \hspace{-2pt} [1\hspace{-2pt}+\hspace{-2pt}o(1)] \label{HC-leq}
 \end{align}
 and
  \begin{align}
 \hspace{-1pt} \mathbb{P}[\hspace{.5pt}G(n, K_n, P_n)\text{ has a \texttt{HamiCycle}.}\hspace{.5pt}] & \hspace{-2pt} \geq \hspace{-2pt}e^{- e^{-\lim\limits_{n \to \infty}{\beta_n}}} \hspace{-2pt}\cdot\hspace{-2pt} [1\hspace{-2pt}-\hspace{-2pt}o(1)].  \label{HC-geq}
 \end{align}
 
 (\ref{HC-leq}) clearly holds from Lemma \ref{cp_rig_er} with $k=2$ and the fact that a necessary condition for a graph to contain a Hamilton cycle is that the minimum degree is at least $2$.
 
 Now we establish (\ref{HC-geq}). From Lemma \ref{lem-only-prove-lnlnn} and the fact that Hamilton cycle containment is a monotone increasing graph property, we can introduce an auxiliary condition $|\beta_n |= O(\ln \ln n)$. We will use Lemmas \ref{cp_rig_er} and \ref{lem:ER:HC} to prove (\ref{HC-geq}). We first show that the conditions of Lemma \ref{cp_rig_er} all hold given the conditions of Theorem \ref{thm:HC} and the auxiliary condition $|\beta_n |= O(\ln \ln n)$. From $|\beta_n |= O(\ln \ln n)$ and (\ref{thm:HC:pe}), (\ref{KnPnnlnn}) still follows here.
As explained before, we have $ \frac{{K_n}^2}{P_n}  =
 o\left( \frac{1}{\ln n} \right)$ and $ \frac{K_n}{P_n}  = o\left( \frac{1}{n\ln n} \right)$ from (\ref{KnPnnlnn}) and $P_n = \omega\big(n (\ln n)^5\big)$.
Then all conditions in Lemma \ref{cp_rig_er} hold. From (\ref{thm:HC:pe}) and (\ref{KnPnnlnn}), the sequence $s_n$ in Lemma \ref{cp_rig_er} satisfies 
 \begin{align}
s_n & =   \frac{\ln  n   + \ln \ln n +
 {\beta_n}}{n} - \frac{\ln  n}{n} \cdot [1 \pm o(1)] \cdot  o\left(\frac{1}{ \ln n}\right) \nonumber \\ & =  \frac{\ln  n     + \ln \ln n +
 {\beta_n}-o(1)}{n}  ,\label{KnPnnlnnres2}
\end{align}
which is used in Lemma \ref{lem:ER:HC} to induce
 \begin{align}
 \lim_{n \to \infty}   \mathbb{P}[\hspace{1pt}G_{ER}(n,s_n)\text{ has a \texttt{HamiCycle}.}\hspace{1pt}] &  = e^{- e^{-\lim\limits_{n \to \infty}{\beta_n}}}. \label{KnPnnlnn3}
 \end{align}
 Then (\ref{HC-geq}) clearly follows from (\ref{KnPnnlnn3}) and Lemma \ref{cp_rig_er}.
 
We have established  Theorem \ref{thm:HC} by showing (\ref{HC-leq}) and (\ref{HC-geq}).

 We now explain that Theorem \ref{thm:HC} still follows if $ \frac{{K_n}^2}{P_n}$ in (\ref{thm:HC:pe}) is replaced by the edge probability denoted by $q_n$; i.e., $q_n = \frac{\ln  n  + \ln \ln n +
 {\beta_n}}{n}$. First, from \cite[Proof of Lemma 14]{ANALCO}, we can also introduce an auxiliary condition $|\beta_n |= O(\ln \ln n)$, which along with $q_n = \frac{\ln  n + \ln \ln n  +
 {\beta_n}}{n}$ yields $q_n = \frac{\ln  n}{n} \cdot [1 \pm o(1)]$. Then
from \cite[Lemma 8]{ZhaoYaganGligor}, it is straightforward to derive $ \frac{{K_n}^2}{P_n}  =  \frac{\ln  n  + \ln \ln n +
 {\beta_n} \pm o(1)}{n} $. Thus Theorem \ref{thm:HC} still follows.
 
 \subsection{Proof of Theorem \ref{thm:PHW}}

We define $F_n(\mathcal{I})$ as
  \begin{align}
\hspace{-1.5pt}F_n(\mathcal{I}) \hspace{-1.5pt} = \hspace{-1.5pt} \begin{cases} \ln n,~\hspace{2.5pt}\text{if $\mathcal{I}$ is perfect matching containment}, \\
\ln n + \ln \ln n,~\hspace{2.5pt}\text{if $\mathcal{I}$ is Hamilton cycle containment}, \\
\ln n + (k-1) \ln \ln n,~\hspace{2.5pt}\text{if $\mathcal{I}$ is $k$-connectivity}. \end{cases} \label{eq-FnI}
 \end{align}
Recalling $K_n^{-}(\mathcal{I}, \epsilon)$ specified in (\ref{KnI-minus}), we further define $\alpha_n^{-}(\mathcal{I}, \epsilon)$ such that
\begin{align}
 \frac{{[K_n^{-}(\mathcal{I}, \epsilon)]}^2}{P_n} & = \frac{F_n(\mathcal{I})   +
 {\alpha_n^{-}(\mathcal{I}, \epsilon)}}{n}.\label{KnI-minus-alpha}
\end{align}
We now use 
   \begin{align}
 \mathbb{P}[\hspace{1pt}G(n, K_{n}^{-}(\mathcal{I}, \epsilon), P_{n})\text{ has property $\mathcal{I}$.}\hspace{1pt}]  &  \geq \epsilon \label{geqepsilon}
 \end{align}
 to establish for any positive constant $\delta_1 < \epsilon$ that
 \begin{align}
 \frac{{[K_n^{-}(\mathcal{I}, \epsilon)]}^2}{P_n} \hspace{-2pt} \geq  \hspace{-2pt} \frac{F_n(\mathcal{I}) \hspace{-2pt}-\hspace{-2pt}\ln (\hspace{-1pt}-\hspace{-1pt}\ln \delta_1)}{n}\text{ for all $n$ sufficiently large,}  \label{PMalpha-lowerboundsb}
\end{align}
where (\ref{geqepsilon}) holds due to (\ref{KnI-minus}).
 With $\alpha_n^{-}(\mathcal{I}, \epsilon)$ defined in (\ref{KnI-minus-alpha}), (\ref{PMalpha-lowerboundsb}) is equivalent to
\begin{align}
\alpha_n^{-}(\mathcal{I}, \epsilon)  \geq  -\ln (-\ln \delta_1)\text{ for all $n$ sufficiently large.} \label{PMalpha-lowerbound}
\end{align}
By contradiction, if (\ref{PMalpha-lowerbound}) is not true,
 there exists a subsequence $N_i |_{i=1,2,\ldots}$ of $\mathbb{N}$ (the set of 
 all positive integers) such that 
  $\alpha_{N_i}^{-}(\mathcal{I}, \epsilon)  <  -\ln (-\ln \delta_1)$ for $i=1,2,\ldots$. By Lemma \ref{lem:subseq}, there exists a subsequence $M_j |_{j=1,2,\ldots}$ of $N_i |_{i=1,2,\ldots}$ such that $\lim_{j \to \infty} \alpha_{M_j}^{-}(\mathcal{I}, \epsilon) \in [-\infty, \infty]$. From $\alpha_{N_i}  < -\ln (-\ln \delta_1)$ for $i=1,2,\ldots$, we have $\lim_{j \to \infty} \alpha_{M_j}^{-}(\mathcal{I}, \epsilon)\in [-\infty,  -\ln (-\ln \delta_1)]$. Then from Lemma \ref{lem:subsequenc}, it follows that
  \begin{align}
 & \lim_{j \to \infty}   \mathbb{P}[\hspace{1pt}G(M_j, K_{M_j}^{-}(\mathcal{I}, \epsilon), P_{M_j})\text{ has property $\mathcal{I}$.}\hspace{1pt}] \nonumber \\  &  \quad = e^{- e^{-\lim\limits_{j \to \infty}{\alpha_{M_j}^{-}(\mathcal{I}, \epsilon)}}} \leq e^{-e^{\ln (-\ln \delta_1)}}  = \delta_1,\nonumber
 \end{align}
 which contradicts (\ref{geqepsilon}).
Therefore, (\ref{PMalpha-lowerboundsb}) and (\ref{PMalpha-lowerbound}) are established. 

Similar to the above analysis of using (\ref{geqepsilon}) to prove (\ref{PMalpha-lowerboundsb}), we can use
  \begin{align}
 \mathbb{P}[\hspace{1pt}G(n, K_{n}^{-}(\mathcal{I} \hspace{-2pt}- \hspace{-2pt}1, \epsilon), P_{n})\text{ has property $\mathcal{I}$.}\hspace{1pt}]  &  < \epsilon \label{geqepsilon-a}
 \end{align}
 to prove 
for any positive constant $\delta_2 > \epsilon$ that
\begin{align}
\hspace{-2pt}\frac{{[K_n^{-}(\mathcal{I},\hspace{-1.5pt} \epsilon) \hspace{-2pt}-\hspace{-2pt} 1]}^2}{P_n}  \hspace{-2.5pt}\leq \hspace{-2.5pt} \frac{F_n(\mathcal{I}) \hspace{-2.5pt}-\hspace{-2.5pt}\ln (\hspace{-1pt}-\hspace{-1.5pt}\ln \delta_2)}{n}\text{ \hspace{-1pt}for \hspace{-1.5pt}all \hspace{-1.5pt}$n$ \hspace{-1.5pt}sufficiently \hspace{-1.5pt}large,}  \label{PMalpha-lowerboundsb-a}
\end{align}
use 
   \begin{align}
 \mathbb{P}[\hspace{1pt}G(n, K_{n}^{+}(\mathcal{I}, \epsilon), P_{n})\text{ has property $\mathcal{I}$.}\hspace{1pt}]  &  \geq 1-\epsilon \label{geqepsilon-b}
 \end{align}
 to prove
 for any positive constant $\delta_3 < 1 - \epsilon$ that
\begin{align}
 \frac{{[K_n^{+}(\mathcal{I}, \epsilon)]}^2}{P_n} \hspace{-2pt} \geq  \hspace{-2pt} \frac{F_n(\mathcal{I})\hspace{-2pt} -\hspace{-2pt}\ln (\hspace{-1pt}-\hspace{-1pt}\ln \delta_3)}{n}\text{ for all $n$ sufficiently large,}  \label{PMalpha-lowerboundsb-b}
\end{align}
and use
   \begin{align}
 \mathbb{P}[\hspace{1pt}G(n, K_{n}^{+}(\mathcal{I} - 1, \epsilon), P_{n})\text{ has property $\mathcal{I}$.}\hspace{1pt}]  &  < 1-\epsilon \label{geqepsilon-c}
 \end{align}
  to prove
for any positive constant $\delta_4 > 1 - \epsilon$ that
\begin{align}
\hspace{-2pt}\frac{{[K_n^{+}(\mathcal{I},\hspace{-1.5pt} \epsilon) \hspace{-2pt}-\hspace{-2pt} 1]}^2}{P_n}  \hspace{-2.5pt}\leq \hspace{-2.5pt} \frac{F_n(\mathcal{I}) \hspace{-2.5pt}-\hspace{-2.5pt}\ln (\hspace{-1pt}-\hspace{-1.5pt}\ln \delta_4)}{n}\text{ \hspace{-1pt}for \hspace{-1.5pt}all \hspace{-1.5pt}$n$ \hspace{-1.5pt}sufficiently \hspace{-1.5pt}large,}  \label{PMalpha-lowerboundsb-c}
\end{align}
Considering that the proofs of (\ref{PMalpha-lowerboundsb-a}) (\ref{PMalpha-lowerboundsb-b}) and (\ref{PMalpha-lowerboundsb-c}) are very similar to that of (\ref{PMalpha-lowerboundsb}), we omit the details here due to space limitation. Note that (\ref{geqepsilon-a}) holds from (\ref{KnI-minus}), while (\ref{geqepsilon-b}) and (\ref{geqepsilon-c}) follow from (\ref{KnI-plus}).

From (\ref{PMalpha-lowerboundsb}) and (\ref{PMalpha-lowerboundsb-a}), it follows that
\begin{align}
\hspace{-1.5pt} \sqrt{\frac{P_n[F_n(\mathcal{I}) -\ln (-\ln \delta_1)]}{n}} &\hspace{-1.5pt} \leq \hspace{-1.5pt}K_n^{-}(\mathcal{I}, \epsilon)   \nonumber \\
&  \hspace{-1.5pt}\leq \hspace{-1.5pt}\sqrt{\frac{P_n[F_n(\mathcal{I}) -\ln (-\ln \delta_2)]}{n}}\hspace{-1.5pt}+\hspace{-1.5pt}1. \label{KnI-minus-alpha-plus-x}
\end{align}
From (\ref{PMalpha-lowerboundsb-b}) and (\ref{PMalpha-lowerboundsb-c}), it follows that
\begin{align}
\hspace{-1.5pt}\sqrt{\frac{P_n[F_n(\mathcal{I}) -\ln (-\ln \delta_3)]}{n}} &  \hspace{-1.5pt} \leq \hspace{-1.5pt}K_n^{+}(\mathcal{I}, \epsilon)   \nonumber \\
&   \hspace{-1.5pt} \leq\hspace{-1.5pt} \sqrt{\frac{P_n[F_n(\mathcal{I}) -\ln (-\ln \delta_4)]}{n}}\hspace{-1.5pt}+\hspace{-1.5pt}1.  \label{KnI-minus-alpha-plus-y}
\end{align}
With the phase transition width $ d_n(\mathcal{I},  \epsilon)$ defined in (\ref{phase-transition-width}), we obtain from (\ref{KnI-minus-alpha-plus-x}) and (\ref{KnI-minus-alpha-plus-y}) that
\begin{align}
 & d_n(\mathcal{I},  \epsilon) +1   \nonumber \\
&  \geq  \sqrt{\frac{P_n[F_n(\mathcal{I}) -\ln (-\ln \delta_3)]}{n}}  -  \sqrt{\frac{P_n[F_n(\mathcal{I}) -\ln (-\ln \delta_2)]}{n}}\label{dnlower}
\end{align}
%
and
\begin{align}
&d_n(\mathcal{I},  \epsilon)  - 1   \nonumber \\
&  \leq  \sqrt{\frac{P_n[F_n(\mathcal{I}) -\ln (-\ln \delta_4)]}{n}}  -  \sqrt{\frac{P_n[F_n(\mathcal{I}) -\ln (-\ln \delta_1)]}{n}}  \label{dnupper}
\end{align}
To use (\ref{dnlower}) and (\ref{dnupper}), we compute
for constants $c_1$ and $c_2$ that 
\begin{align}
& \sqrt{\frac{P_n(F_n(\mathcal{I})+c_1)}{n}}  - \sqrt{\frac{P_n(F_n(\mathcal{I})+c_2)}{n}}  \vspace{-2pt} \nonumber \\
& \quad = \left[\frac{P_n(F_n(\mathcal{I})+c_1)}{n} - \frac{P_n(F_n(\mathcal{I})+c_2)}{n}\right]   \vspace{-2pt} \nonumber \\
& \quad\quad\times  \left[ \sqrt{\frac{P_n(F_n(\mathcal{I})+c_1)}{n}}  + \sqrt{\frac{P_n(F_n(\mathcal{I})+c_2)}{n}}\hspace{2pt}\right]^{-1}  \vspace{-2pt} \nonumber \\
&\quad \sim \frac{P_n(c_1-c_2)}{n} \times \left(2\sqrt{\frac{P_n F_n(\mathcal{I})}{n}}\hspace{2pt}\right)^{-1} \vspace{-2pt}  \nonumber \\
&\quad \sim \frac{1}{2}  (c_1-c_2) \sqrt{\frac{P_n}{n \ln n}}   . \label{dnlowerupper}
\end{align}
where the last step uses
$F_n(\mathcal{I}) \sim \ln n$ from (\ref{eq-FnI}) (an asymptotic equivalence $x_n \sim y_n$ means \vspace{.5pt} $\lim_{n \to \infty} \frac{x_n}{y_n} = 1$). 
 With R.H.S. meaning the right hand side, we get from (\ref{dnlowerupper}) that \\a) if $P_n = o(n\ln n)$, then $\text{R.H.S. of (\ref{dnlower})}=o(1)$ and $\text{R.H.S. of (\ref{dnupper})}=o(1)$, \\b) if $P_n = \Theta(n\ln n)$, then $\text{R.H.S. of (\ref{dnlower})}=\Theta(1)$ and $\text{R.H.S. of (\ref{dnupper})}=\Theta(1)$, and \\c) if $P_n = \omega(n\ln n)$, then $\text{R.H.S. of (\ref{dnlower})}=\omega(1)$ and $\text{R.H.S. of (\ref{dnupper})}=\omega(1)$.

In view that the results in Lemma \ref{thm:ZhaoCDCk} on $k$-connectivity has the condition $P_n = \Omega(n)$, and that the results in 
Theorems \ref{thm:PM} and  \ref{thm:HC} on perfect matching containment and Hamilton cycle containment respectively both rely on the condition $P_n = \omega\big(n (\ln n)^5\big)$, we obtain the desired results (i)--(iv) from (\ref{dnlower}) (\ref{dnupper}), the above a), b) and c), and $d_n(\mathcal{I}, \epsilon)\geq 0$.  

\section{Related Work} \label{related}

 For connectivity in $G_u(n,P_n,K_n)$, Blackburn and Gerke \cite{r1}, and Ya\u{g}an and Makowski
\cite{yagan} obtain different granularities of zero--one laws, and 
 Rybarczyk
\cite{ryb3} establishes the asymptotically exact probability result. For $k$-connectivity,  Rybarczyk \cite{zz}
implicitly shows a zero--one law, and we \cite{ZhaoCDC} derive the asymptotically exact probability, as used in Lemma \ref{thm:ZhaoCDCk} of this current paper. For perfect matching containment, Bloznelis and {\L}uczak \cite{Perfectmatchings} give the asymptotically exact probability result, 
but their result after a rewriting applies to a different set of conditions on $P_n$ compared with our Theorem \ref{thm:PM} which is for $P_n = \omega\big(n (\ln n)^5\big)$: they consider $P_n = \Omega\big(n (\ln n)^{-1}\big)$ and $P_n = o\big(n (\ln n)^{-\frac{3}{5}}\big)$ instead. 
 In terms of Hamilton cycle containment, Nikoletseas \emph{et al. } \cite{NikoletseasHM} proves that 
 $G(n, K_n, P_n)$ under $K_n \geq 2$ has a
    Hamilton cycle with high probability 
     if it holds for some constant $\delta>0$ that $n \geq (1+\delta) \binom{P_n}{K_n} \ln \binom{P_n}{K_n} $, which implies that $P_n$ is much smaller than $n$ ($P_n = O(\sqrt{n}\hspace{1.5pt})$ given $K_n \geq 2$,
    $P_n = O(\sqrt[3]{n}\hspace{1.5pt})$ if $K_n \geq 3$, $P_n = O(\sqrt[4]{n}\hspace{1.5pt})$ if $K_n \geq 4$, etc.). Different from the result of Nikoletseas \emph{et al. } \cite{NikoletseasHM}, our Theorem \ref{thm:HC} is for $P_n = \omega\big(n (\ln n)^5\big)$. Furthermore, Theorem \ref{thm:HC} presents the asymptotically exact probability, whereas Nikoletseas \emph{et al. } \cite{NikoletseasHM} only derive conditions for $G_u(n,P_n,K_n)$ to have a Hamilton cycle with high probability. They do not provide conditions for $G_u(n,P_n,K_n)$ to have no Hamilton cycle with high probability, or to have a Hamilton cycle with an asymptotic probability in $(0,1)$.

A graph model related to uniform random intersection graphs is the so-called \emph{binomial random intersection graph model}: each item in a pool is assigned to each node independently with the same probability, and two nodes establish an undirected edge upon sharing at least one item. Note that here the number of items on each node follows a binomial distribution. This graph model has also been studied in the literature as noted below. 
 For connectivity, Rybarczyk presents a zero--one law
\cite{zz} and later obtains the stronger result of the asymptotically exact probability \cite{2013arXiv1301.0466R}. For $k$-connectivity, 
she establishes zero--one laws \cite{zz,2013arXiv1301.0466R}, and we compute the asymptotically exact probability \cite{ZhaoCDC}. For perfect matching containment, Rybarczyk provides a zero--one law
\cite{zz} and derives the asymptotically exact probability \cite{2013arXiv1301.0466R}.
 For Hamilton cycle containment, Efthymioua and Spirakis \cite{EfthymiouaHM}, and Rybarczyk \cite{zz,2013arXiv1301.0466R} show zero--one laws.

\section{Conclusion}
\label{sec:Conclusion}

In a uniform random intersection graph,
for perfect matching containment and Hamilton cycle containment, we derive the asymptotically exact probabilities and the phase transition widths. In addition, for $k$-connectivity, we use the asymptotically exact probability result in our prior work \cite{ZhaoCDC} to compute the phase transition width.



 \renewcommand\baselinestretch{.95}

 \renewcommand\baselinestretch{1}

 \appendix
 

\subsection{Proof of Lemma \ref{thm:ZhaoCDCk} in Section \ref{sec:factlem}}  \label{thm:ZhaoCDCk:lemprf}

From Lemma \ref{lem-only-prove-lnlnn} and the fact that both $k$-connectivity and the property of minimum degree being at least $k$ are monotone increasing, we can introduce an auxiliary condition $|\alpha_n |= O(\ln \ln n)$ in proving Lemma \ref{thm:ZhaoCDCk}. 
From $|\gamma_n |= O(\ln \ln n)$ and (\ref{thm:PM:pe}), it follows that $ \frac{{K_n}^2}{P_n}   \sim \frac{\ln  n}{n} $, which along with $P_n = \Omega(n)$ implies $K_n = \Omega \big(\sqrt{\ln n}\hspace{2pt}\big)$. Under $K_n = \Omega \big(\sqrt{\ln n}\hspace{2pt}\big)$ and (\ref{thm:pe:kgencdc}) with $|\gamma_n |= O(\ln \ln n)$, the result (\ref{thm:pe:kgencdcres}) follows from our work \cite[Theorem 1]{ZhaoCDC}. Then in view of Lemma \ref{lem-only-prove-lnlnn}, we have completed proving Lemma \ref{thm:ZhaoCDCk}.
 
 \subsection{Proof of Lemma \ref{cp_rig_er} in Section \ref{sec:factlem}}  \label{cp_rig_er:lemprf}

 From our work \cite[Lemmas 3 and 5]{ZhaoCDC}, with $t_n$ defined by $t_n =  \frac{K_n}{P_n}
 \big(1 - \sqrt{\frac{3\ln
n}{K_n }}\hspace{2pt}\big)$, if 
$t_n = o\left( \frac{1}{n\ln n} \right)$ and ${t_n}^2
P_n =
 o\left( \frac{1}{\ln n} \right)$, then there exists $s_n =
{t_n}^2 P_n \cdot \left[1-
 o\left(\frac{1}{ \ln n}\right)\right]$ such that (\ref{cp_res_rig_er}) holds. Hence, the proof of Lemma \ref{cp_rig_er} will be completed once we show 
i) $t_n = o\left( \frac{1}{n\ln n} \right)$, ii) ${t_n}^2
P_n =
 o\left( \frac{1}{\ln n} \right)$ \vspace{1pt} and that iii) $s_n $ being $
{t_n}^2 P_n \cdot \left[1-
 o\left(\frac{1}{ \ln n}\right)\right]$ can be written as $ \frac{{K_n}^2}{P_n} \cdot \left[1-
 o\left(\frac{1}{ \ln n}\right)\right]$. Given conditions $K_n = \omega\big((\ln n)^3\big)$, $ \frac{K_n}{P_n} = o\left( \frac{1}{n\ln n} \right)$\vspace{1pt} and $ \frac{{K_n}^2}{P_n}  =
 o\left( \frac{1}{\ln n} \right)$ in Lemma \ref{cp_rig_er}, and using $t_n =  \frac{K_n}{P_n}
 \big(1 - \sqrt{\frac{3\ln
n}{K_n }}\hspace{2pt}\big)$, we clearly obtain i), ii) and iii) above.



\end{document}